\documentclass[letterpaper,twoside,twocolumn,english,aps,prl]{revtex4}
\usepackage[T1]{fontenc}
\usepackage[latin9]{inputenc}
\usepackage{verbatim}
\usepackage{amsmath}
\usepackage{amssymb}
\usepackage{esint}

\makeatletter

\@ifundefined{textcolor}{}
{%
 \definecolor{BLACK}{gray}{0}
 \definecolor{WHITE}{gray}{1}
 \definecolor{RED}{rgb}{1,0,0}
 \definecolor{GREEN}{rgb}{0,1,0}
 \definecolor{BLUE}{rgb}{0,0,1}
 \definecolor{CYAN}{cmyk}{1,0,0,0}
 \definecolor{MAGENTA}{cmyk}{0,1,0,0}
 \definecolor{YELLOW}{cmyk}{0,0,1,0}
}

\makeatother

\usepackage{babel}
\begin{document}
\global\long\def\ket#1{\left|#1\right\rangle }

\global\long\def\bra#1{\left\langle #1\right|}

\global\long\def\braket#1#2{\left\langle #1\left|\vphantom{#1}#2\right.\right\rangle }

\global\long\def\ketbra#1#2{\left|#1\vphantom{#2}\right\rangle \left\langle \vphantom{#1}#2\right|}

\global\long\def\braOket#1#2#3{\left\langle #1\left|\vphantom{#1#3}#2\right|#3\right\rangle }

\global\long\def\mc#1{\mathcal{#1}}

\global\long\def\nrm#1{\left\Vert #1\right\Vert }

\title{Time-energy tradeoff in unambiguous state discrimination POVM}

\author{Raam Uzdin}

\author{Omri Gat}

\address{Racah Institute of Physics, The Hebrew University, Jerusalem, Israel.}

\email{raam@mail.huji.ac.il}

\selectlanguage{english}%
\begin{abstract}
Unambiguous non-orthogonal state discrimination has fundamental importance
in quantum information and quantum cryptography. The discrimination
is carried out by POVM generalized measurements. For this process,
we find a tradeoff between the needed energy resources and the evolution
time, and express it in terms of action-like cost inequality. We find
the realization that minimizes this action-like cost and show that
in this case the cost is determined by the maximal population transfer
from the system to the ancilla needed for the POVM implementation. 
\end{abstract}
\maketitle
Non-stationary processes in quantum mechanics involve an intrinsic
energy cost that is inversely proportional to the time duration of
the process. The exact relation, however, depends on the details of
the process. Anandan and Aharonov \cite{Aharonov-Anandan} established
a relation between the energy variance of the Hamiltonian and the
rotation time of a state in Hilbert space (e.g spin $1/2$ flips).
A more general relation between the norm action of the Hamiltonian
and the evolution operator of a process was derived by Lidar, Zanardi
and Khodjasteh \cite{Lidar}. A similar result holds for systems with
absorption (loss of probability) \cite{NU resources}. In this paper
we derive the time-energy constraint for the fundamental quantum process
of unambiguous states discrimination (USD \cite{Barnett Book,Jaeger Book,Bergou SD rev,Chefles}).

In an ideal USD, a system is prepared randomly in one of a set of
a priory known \textit{non-orthogonal} states. The goal is to detect
the system's state with zero error probability. This problem has led
to a deeper understanding of what information can be extracted from
a quantum system and at what probabilistic cost. From the practical
point of view USD can be used for secure quantum communication \cite{Bennett}
and for entanglement distillation \cite{Chefles,SVD USD}. 

Although USD of non-orthogonal states cannot be realized by standard
(von Neumann) projective measurements (without dilating the Hilbert
space), it can be implemented without errors by a generalized measurement
known as POVM (positive-operator valued measure\cite{Peres,Nielsen}).
Unfortunately, POVM implementation of USD inherently involve some
non-zero probability of obtaining an inconclusive result, from which
the input state cannot be inferred. The inconclusive result probability
depends on the degree of non-orthogonality of the input state. 

In this work we consider the unitary embedding POVM scheme \cite{Hutnner,Bergou embed 2000,Bergou embed 2006,Uwe neum brach,Williams book}
where the system subspace is coupled to ancillary levels by a unitary
evolution and the state detection is carried out by standard von Neuman
measurement in the system subspace only. Here, we quantify the minimal
time-energy resources associated with the unitary evolution in this
scheme. According to the Neumark dilation theorem \cite{Peres}, a
POVM can also be implemented directly as a von Neumann measurement
in a larger Hilbert space (without a pre-measurement evolution). However,
if we require that the conclusive measurement results appear exclusively
in the original system subspace as in the unitary embedding scheme,
a unitary evolution must be applied after the measurement. We show
that the cost of the post-measurement ``information concentrating''
unitary is exactly equal to the unitary evolution cost in the unitary
embedding scheme described above. In the unitary embedding scheme
we find that the minimal time-energy cost is determined by the maximal
population transfer from the system subspace to the ancilla subspace.

Note that previous studies about resources of unitary evolution such
as \cite{Lidar} cannot be immediately applied to the unitary embedding
scheme studied here, since the USD process provides only partial information
on the unitary evolution operator. Finally we comment that USD requires
other resources beside energy. For example the entanglement cost of
a general rank-one POVM embedding was studied in \cite{Popescu}.

\textbf{\textit{USD POVM and lossy evolution}}$-$In a POVM measurement
each measurement result '$i$' is associated with a positive operator
$F_{i}$. Given a density matrix $\rho$, the probability to get the
result '$i$' is $p_{i}=\text{tr}(\rho F_{i})$. For a USD of $N$
non-orthogonal states in a Hilbert space of dimension $N$, the $\{F_{i}\}_{i=1}^{N}$
rank-one operators are constructed from the bi-orthogonal basis \cite{equive RU}.
An additional operator (that is typically not rank one) is defined
as $F_{N+1}=I-\sum_{n=1}^{N}F_{n}$, and it describes the inconclusive
result. Huttner et al.\ \cite{Hutnner} first suggested and experimentally
demonstrated that USD POVM can be implemented by a lossy evolution.
Recently, the equivalence between USD POVM and lossy evolution was
further studied \cite{equive RU}. An $N$-level state evolving under
a lossy evolution operator $K$ satisfies: $\ket{\psi_{\text{final}}}=K\ket{\psi_{\text{initial}}}$
where $K\in C^{N\times N}$ is not a unitary operator. Such an evolution
does not conserve the angle between states. This fact can be exploited
in order to transform the non-orthogonal states at the input to orthogonal
states at the output. Once the states are orthogonal they can be discriminated
without errors using a regular projective measurement. The inherent
loses in the system make the detection probability smaller than one.
This probability loss is mathematically equivalent to the inconclusive
result in the POVM formalism. In fact any USD POVM $\{F_{i}\}_{i=1}^{N+1}$
can be associated with a lossy evolution operator and vice versa \cite{equive RU}.
Apart from the practical value of this equivalence for USD realization,
the lossy evolution approach has theoretical merits as well. In particular,
it was shown in \cite{SVD USD} that the singular values of the lossy
evolution operator capture the essence of USD and can be used to reveal
interesting insights into multiple USD. 

\textbf{\textit{Embedding of a lossy evolution}}$-$There are two
different ways of implementing a lossy evolution as defined above.
The first is to find a system described by some effective non-Hermitian
Hamiltonian \cite{Nim book} that includes losses (e.g. optics with
non-negligible absorption). The other way is to consider the evolution
of a closed unitary system and measure the evolution outcomes only
in a subsystem. Assume that initially the total probability of finding
particle in a subsytem is unity. After a unitary evolution of the
whole system is completed the total probability in the subsystem is
typically less than one. From the point of view of the subsystem the
evolution is lossy. As discussed later, any lossy evolution can be
embedded in a larger Hilbert space unitary evolution. We refer to
this second implimentation as ``unitary embedding''. The two alternatives
are closely related. Yet, in some systems the effective Hamiltonian
is much more accessible (e.g. in optics with absorption) and in other
systems the unitary of the whole system is easier to work with (e.g.
a few qubits coupled to an ancilla qubit). The resources needed to
implement the first scheme were studied in \cite{NU resources}. In
this work we focus on the embedding scheme. Let $K$ be a lossy evolution
operator $K\in\mathbb{C}^{N\times N}$ that implements some desired
USD POVM . We wish to implement $K\ket{\psi}$ by embedding $K$ and
$\ket{\psi}$ in a larger Hilbert space governed by a unitary evolution.
In this scheme the embedding unitary evolution $U$ satisfies:
\begin{equation}
U\begin{pmatrix}\ket{\psi}\\
0
\end{pmatrix}=\left(\begin{array}{c|c}
K & B\\
\hline C & D
\end{array}\right)\begin{pmatrix}\ket{\psi}\\
0
\end{pmatrix}=\begin{pmatrix}K\ket{\psi}\\
C\ket{\psi}
\end{pmatrix}\begin{array}{c}
\}\text{System}\\
\}\text{Ancilla}
\end{array},\label{eq: U embed}
\end{equation}
where we measure only the first $N$ levels (the system). For example,
the ancilla levels can describe unmeasured waveguides or atomic levels
(see examples in \cite{opt and sold disc}). Although the motivation
for this work comes from the USD problem the finding presented here
are applicable to any embedding of lossy evolution operator. For example,
it can be used for entanglement distillation \cite{SVD USD}.

There are many different degrees of freedom in choosing $B$,$C,$
and $D$. $K$ is determined (up to a multiplication by unitary matrix
from the left) by the desired USD POVM. Given only $K$, our goal
is to find the choice that minimizes the time-energy resources defined
in the next section. Furthermore we want to obtain an explicit expression
for the time-energy cost in this case.

\textbf{\textit{Resources and norm action}}$-$Let $U$ be some unitary
evolution operator generated by a Hamiltonian $H$ so that $i\frac{d}{dt}U=HU$.
The construction of $H$ requires some physical resources like magnetic
field or coupling to laser radiation. To quantify the resources, the
Hamiltonian must be mapped to a scalar. Here, we use unitarily invariant
matrix norms for this purpose. There are three main reasons for using
these norms as measures of resources. First, it is natural to demand
that a measure of resources will satisfy the defining properties of
a norm \cite{Horn}. Second, unitary invariance insures that the resources
do not depend on the basis in which the Hamiltonian is expressed.
Third, we shall use an important relation between the evolution operator
and unitarily invariant norms of the Hamiltonian \cite{Lidar}. The
time integral over the Hamiltonian norm is called the ``norm action''.
From the result of \cite{Lidar} it follows that the norm action (LHS
of (\ref{eq: Lidar})) of the Hamiltonian is bounded from below by
$U$ in the following way:
\begin{equation}
\int_{0}^{T}\nrm{H(t)}dt\ge\nrm{\ln U(T)},\label{eq: Lidar}
\end{equation}
where in the $\ln$ the angles are in the branch $(-\pi,\pi]$, and
$\nrm{\cdot}$ may refer to any unitarily invariant matrix norm. When
the Hamiltonian is time-independent the norm action integral on the
LHS of (\ref{eq: Lidar}) reduces to $\text{time}\times\text{energy}$
and the inequality becomes an equality. In this case $U=e^{-iH_{0}T}$
where $H_{0}$ is the generating Hamiltonian and $T$ is the duration
of the evolution. At this point it is still not clear what is the
embedding $U$ that yields a minimal norm action, but it is clear
that this embedding must be generated by some time-independent Hamiltonian
$H_{0}$ that can be obtained from $U$ (with the assumption made
above on the branch of the $\ln$ function). Action-like quantities
have been used before to analyze quantum evolution \cite{Zurek action}.
Unless stated otherwise, in this work, we shall use the spectral norm
\cite{Horn}. It has a clear physical interpretation \cite{NU speed,NU resources}
and it leads to compact and comprehensible results. The spectral norm
of a matrix $A\in\mathbb{C}^{N\times N}$ is the largest singular
value of $A$. $ $The singular values $s_{i}$ of $A$ are: $\{s_{i}\}=\sqrt{\text{eigenvalues}[A^{\dagger}A]},$
and therefore: $\nrm A=s_{\text{max}}=\max(\sqrt{\text{eigenvalues}[A^{\dagger}A]}).$ 

The fact that lossy systems cannot amplify the amplitude of any state,
manifests itself in the condition $\nrm K\le1$ \cite{NU resources}.
If $\nrm K=1$ the system is called ``marginally passive'' \cite{NU resources}.
Finally, notice that for \textit{Hermitian} \textit{Hamiltonians}
the spectral norm is equal to the largest absolute valued energy.

\textbf{\textit{Decomposition of the unitary embedding}}$\mathit{-}$Our
strategy of finding the USD-generating Hamiltonian with the minimal
possible norm action is the following. We start by finding an explicit
expression for the norm action associated with a specific embedding
choice of a general UDS task. Next, we show that this choice is the
norm action minimizer among all possible unitary embeddings implementing
the same USD task.

Until stated otherwise we shall assume that the number of ancilla
levels and the number of system level is equal. This is enough to
allow the embedding of the most general lossy evolution operator.
Using the polar decomposition of blocks $K$ and $D$, any unitary
$U$ can be written as product of block diagonal unitary $V$ and
positive diagonal block unitary $W$:
\begin{eqnarray}
U & = & VW\\
V=\begin{pmatrix}u_{s} & 0\\
0 & u_{a}
\end{pmatrix} & , & W=\begin{pmatrix}\mc K & \mc B\\
\mc C & \mc D
\end{pmatrix}.\label{eq: Ubd UKD}
\end{eqnarray}
$u_{s}$ and $u_{a}$ are unitaries which operate on the system and
ancilla space respectively. $\mc K$ and $\mc D$ are positive matrices.
For $\mathcal{K},\mc D>0$ the unitarity constraints and the SVD of
each block lead to the following general form:

\begin{equation}
W=\left(\begin{array}{c|c}
u_{\mathcal{K}}\cos\Theta u_{\mathcal{K}}^{\dagger} & -iu_{\mc K}\sin\Theta u_{\mathcal{D}}^{\dagger}\\
\hline -iu_{\mathcal{D}}\sin\Theta u_{\mc K}^{\dagger} & u_{\mc D}\cos\Theta u_{\mc D}^{\dagger}
\end{array}\right)\label{eq: Upos}
\end{equation}
where $\Theta$ is a positive diagonal matrix satisfying: 
\begin{equation}
0\le\Theta_{ii}\le\pi/2.
\end{equation}
 The $u_{\mc K}$ and $u_{\mc D}$ are unitaries whose column vectors
are the orthogonal eigenstates of the positive $\mathcal{K}$ and
$\mc D$ respectively. Each of the blocks is now written in terms
of its singular vectors and therefore the diagonal matrices $\cos\Theta$
and $\sin\Theta$ contain the singular values of blocks $\mathcal{K},\mc D$
and of blocks $\mathcal{B},\mc C$ respectively (or alternatively
K,D and B,C). The time-independent Hamiltonian that generates $W$
is given by:
\begin{equation}
H_{W}=H_{\text{opt}}=\frac{1}{T}\left(\begin{array}{c|c}
0 & u_{\mc K}\Theta u_{D}^{\dagger}\\
\hline u_{\mathcal{D}}\Theta u_{K}^{\dagger} & 0
\end{array}\right).\label{eq: Hopt}
\end{equation}
We write $H_{\text{opt}}$ since later on we show that $H_{W}$ is
the most efficient Hamiltonian that implements the desired USD characterized
by $K$. A similar Hamiltonian has been used before in \cite{Williams book}
for probabilistic evolution and for POVM embedding in \cite{Bergou embed 2006}.
Here however we focus on the resources of embedding. Furthermore,
in our scheme it is critical that $0\le\Theta_{ii}\le\pi/2$ so that
$\mc K$ and $\mc D$ are positive. As will be explained later this
is necessary for optimality. By inspecting $H^{\ensuremath{\dagger}}H$
it is easy to verify that:
\begin{equation}
\nrm{H_{\text{opt}}}T=\max(\Theta_{ii})=\arcsin(\nrm B).
\end{equation}
Or in terms of the singular values of $K$ which is directly determined
by the USD POVM:
\begin{eqnarray}
\intop_{0}^{T}\nrm{H_{\text{opt}}}dt & = & \arcsin(\sqrt{1-s_{\text{min}}^{2}})\label{eq: opt norm action}\\
 & = & \arcsin(\sqrt{1-1/\nrm{K^{-1}}^{2}}).
\end{eqnarray}
Note that $K$ can be replaced by $\mathcal{K}$ as they have the
same singular values. The argument of the $\arcsin$ has a clear physical
meaning. It is the maximal fraction of the population that $W$ can
transfer from the system to the ancilla. 

\textbf{\textit{$\mathbf{W}$ requires lowest possible Hamiltonian
resources$-$}}The goal of this section is to show that $W$ requires
the minimal norm action for the given USD task. Let us try to better
understand the relation between the Hamiltonian $H$, $U$ and $\mc K$.
An input state $\ket{\psi_{\text{in}}}$ is transformed by $W$ according
to (\ref{eq: U embed}). $W$ rotates this vector in Hilbert space.
From the overlap of the initial and final state we can obtain the
rotation angle in Hilbert space:
\begin{equation}
\cos\Omega=\left|\braket{\psi_{\text{in}}}{\psi_{\text{out}}}\right|=\left|\braOket{\psi_{\text{in}}}{\mc K}{\psi_{\text{in}}}\right|.\label{eq: cos Om K}
\end{equation}
Since $\mc K$ is positive, the maximal angle is obtained for the
singular vector $\ket{\psi_{min}}$ associated with the minimal singular
vector. Using $\ket{\psi_{\text{in}}}=\ket{\psi_{min}}$ in (\ref{eq: cos Om K})
we get $\cos\Omega_{\text{max},\mc K}=s_{\text{min}}$, or:
\begin{equation}
\Omega_{\text{max},\mc K}=\arcsin\sqrt{1-s_{\text{min}}^{2}},\label{eq: omega max}
\end{equation}
which is exactly equal to (\ref{eq: opt norm action}). Furthermore,
from the Hamiltonian variance \cite{Aharonov-Anandan,NU speed} one
can show that:
\begin{equation}
\Omega\le\int\left|\frac{d\Omega}{dt}\right|dt\le\int\nrm Hdt\label{eq: speed bound omega}
\end{equation}
Using (\ref{eq: omega max}) and (\ref{eq: speed bound omega}) we
get:
\begin{equation}
\int\nrm Hdt\ge\arcsin\sqrt{1-s_{\text{min}}^{2}}.\label{eq: SP norm action}
\end{equation}
However for $\mathcal{K},\mc D>0$ and time-independent Hamiltonian,
we have already shown that there is an equality (\ref{eq: opt norm action}):
$\nrm{H_{\text{opt}}}T=\arcsin\sqrt{1-s_{\text{min}}^{2}}$. This
provides a very intuitive picture of our claim. The needed resources
in this embedding are determined by the state that experiences the
largest population transfer to the ancilla.

To achieve the goal of the section we will show that when applying
an extra block diagonal unitary $V$, that $\ket{\psi_{\text{min}}}$
leads to a larger rotation in Hilbert space compared to the previous
case (and consequently more norm action resources are needed). We
repeat (\ref{eq: cos Om K}) but this time add a unitary $u_{s}$
that operates on the system subspace and obtain:
\begin{eqnarray}
\cos\Omega_{\text{new}} & = & \left|\braOket{\psi_{\text{min}}}{u_{s}\mc K}{\psi_{\text{min}}}\right|\nonumber \\
 & = & \left|\braOket{\psi_{\text{min}}}{u_{s}}{\psi_{\text{min}}}\right|s_{\text{min}}\le s_{\text{min}}.
\end{eqnarray}
Hence, $\Omega_{\text{new}}>\Omega_{\max,\mc K}.$ Using inequality
(\ref{eq: speed bound omega}) once again we get:
\begin{equation}
\int\nrm{H_{\text{new}}}dt\ge\Omega_{\text{new}}\ge\arcsin\sqrt{1-s_{\text{min}}^{2}},
\end{equation}
or stated in a different way:
\begin{equation}
\int\nrm{H_{\text{new}}}dt\ge H_{W}T.
\end{equation}
Repeating this for states that populate only the ancilla and using
the fact that block $\mc K$ and $\mc D$ have the same $s_{\text{min}}$,
we obtain that any block diagonal rotation of the form $V$ (\ref{eq: Ubd UKD})
only increases the Hamiltonian resources with respect to the $H_{\text{opt}}$
that generates $W$. The same claim can be proved for the Hilbert-Schmidt
norm ($\nrm A_{\text{HS}}=\sqrt{\sum_{ij}\left|A_{ij}\right|^{2}}=\sqrt{\text{tr}(A^{\dagger}A)}=\sqrt{\sum s_{i}^{2}}$).
The optimal Hamiltonian is the same but in expression (\ref{eq: opt norm action})
for the norm action the RHS is replaced by $\sqrt{2\sum_{i=1}^{N}(\arcsin\sqrt{1-s_{i}^{2}})^{2}}$
. 

\textbf{\textit{The dimension of the ancilla}}\textbf{$-$}In principle
the ancilla dimension (the number of levels) $N_{a}$ does not have
to be equal to the dimension of the system $N_{s}$. Let us start
with the$N_{a}=N_{s}$ case and see that $N_{a}$ can be changed without
any effect on the norm action as long as it still implements the same
$\mc K$. The off diagonal blocks of the Hamiltonian (\ref{eq: Hopt})
have dimensions $N_{a}\times N_{a}$. However if $M$ singular values
of $K$ are equal to one the upper right block will contain M zero
columns and the bottom left block will contain M zero rows. Cropping
out the zeros rows and columns the new dimension of the reduced unitary
is $[2N_{s}-M]\times[2N_{s}-M]$ which means that only $N_{s}-M$
ancilla level are needed for the embedding. The converse of this claim
is that is that if $N_{a}=N_{s}-M$ then there are at least $M$ singular
values of $K$ that are equal to one. Note that even if one singular
value is equal to one, the lossy evolution operator $K$ must be marginally
passive. An interesting case is $N_{a}=1$ where $K$ has only one
singular value that is not equal to one. In this special case the
inconclusive result POVM operator has rank one. Consequently, after
an inconclusive result the state of the system contains zero information
on the input state (see Sec. III B of \cite{SVD USD} for an explanation).
The $N_{a}>N_{s}$ case can be analyzed by replacing the states in
$u_{\mc D}$ in (\ref{eq: Hopt}) by orthogonal vectors of dimension
$N_{a}>N_{s}$ (so that $u_{\mc D}$ has $N_{s}$ row $N_{a}$ columns)
. This just add zeros to the singular values of $H_{\text{opt}}$.
Hence, extending the ancilla dimension in the way described above
does not change the norm action with respect to the optimal $N_{a}=N_{s}$
studied in the previous sections.

\textbf{\textit{Relation to Neumark Dilation}}$-$Using tensor product
notation for the $N_{a}=N_{s}$ case, the lossy evolution scheme can
be written as:
\begin{equation}
p_{k}=\text{tr}\{U(\rho_{in}\otimes\rho_{\uparrow})U^{\dagger}(\pi_{k}\otimes\rho_{\uparrow})\},\label{eq: tensor lossy}
\end{equation}
where $\pi_{k}$ are von-Neumann projection operators in the system
subspace and $\rho_{\uparrow}=\left(\begin{array}{cc}
1 & 0\\
0 & 0
\end{array}\right)$. Equation (\ref{eq: tensor lossy}) can also be written as:
\begin{eqnarray}
p_{k} & = & \text{tr}\{(\rho_{in}\otimes\rho_{\uparrow})\Pi_{k}\}.\\
\Pi_{k} & = & U^{\dagger}(\pi_{k}\otimes\rho_{\uparrow})U,
\end{eqnarray}
where $\Pi_{k}$ are projection operators in the total system-ancilla
space. These extended projectors constitute a realization of Neumark
dilation. Despite this intimate relation between Neumark dilation
and lossy evolution there are some significant practical and theoretical
differences. From the theoretical point of view, the lossy evolution
approach can be very useful since USD processes can be analyzed by
studying the properties of $K_{N_{s}\times N_{s}}$ only \cite{SVD USD}.
From the practical and physical point of view, we would like to emphasize
that in the Neumark scheme the number of the measured levels is typically
twice as large (as explained earlier) compared to the lossy evolution
scheme proposed above (where only the system levels are measured).
Furthermore, upon a successful discrimination the state of the system
will populate the ancilla level as well, while in the embedding scheme,
only the system levels are populated when a successful detection takes
place.

In contrast to the minimal norm action found above (\ref{eq: opt norm action}),
it appears that the Neumark scheme requires no norm action since a
regular projective measurement is immediately carried out on the input
states without any prior evolution. The reason for the discordance
in the resources needed for the two schemes stems from the fact that
the information is encoded differently in the two approaches. In the
lossy evolution the discrimination results are contained in the $N$
system levels while in the Neumark approach the information is typically
encapsulated in $2N$ levels. If we want to concentrate the successful
Neumark detection to $N$ levels (as in the lossy evolution scheme)
another unitary must be applied to the system after the measurement
has been completed. After the measurement the density matrix of the
system is $\rho_{\text{after}}=\sum_{k=1}^{N_{s}+N_{a}}p_{k}\Pi_{k}$.
The unitary that will change the first $N_{s}$ elements (that corresponds
to a successful discrimination) to: $\sum_{k=1}^{N_{s}}p_{k}\pi_{k}$
is exactly $U$. This concentrating transformation is defined up to
block diagonal unitary rotation (\ref{eq: Ubd UKD}). However, as
we have shown earlier, the most efficient $U$ is the one in which
the diagonal blocks are positive operators. Hence, the minimal cost
of concentrating the conclusive information to $N$ levels is exactly
equal to the cost of the unitary embedding scheme found above (\ref{eq: opt norm action}). 

\textbf{\textit{An example - USD in atomic system coupled to a laser}}$-$Consider
a three-level atomic system in a time-dependent external electric
field $\varepsilon(t)$ (a laser). The first and second levels are
dipole coupled to the third level, but not coupled to each other.
The Hamiltonian is:
\begin{equation}
H_{0}=\begin{pmatrix}E_{1} & 0 & d_{1}\varepsilon(t)\\
0 & E_{2} & d_{2}\varepsilon(t)\\
d_{1}^{*}\varepsilon(t) & d_{2}^{*}\varepsilon(t) & E_{3}
\end{pmatrix}.
\end{equation}
Where the $d_{i}$ are the dipole coupling coefficients. Setting the
time-dependent (real) electric field to be $\varepsilon(t)=a_{1}\cos[(E_{3}-E_{1})t+\varphi_{1}]+a_{2}\cos[(E_{3}-E_{2})t+\varphi_{2}]$
and applying the rotating wave approximation (RWA) we get:
\begin{equation}
H_{\text{RWA}}=\begin{pmatrix}0 & 0 & A_{1}\\
0 & 0 & A_{2}\\
A_{1}^{*} & A_{2}^{*} & 0
\end{pmatrix}
\end{equation}
where $A_{i}=$$\frac{d_{i}}{2}a_{i}e^{-i\varphi_{i}}$. $H_{\text{RWA}}$
has the form of $H_{\text{opt}}$ (\ref{eq: Hopt}) so the final result
will be expressed in terms of the equality (\ref{eq: opt norm action})
rather than (\ref{eq: SP norm action}). We shall use levels one and
two as the ``system'' levels, and the third level will be used as
an ancilla level. One should keep in mind that the rotated wave function
is related to the actual state via: $\ket{\psi}=\exp[-i\:\text{diag}\{E_{1},E_{2},E_{3}\}]\ket{\psi_{\text{RWA}}}$.
However, this rotation has a block diagonal structure with respect
to the system and the ancilla and therefore this transformation will
add a trivial rotation to the subspace of the system but the orthogonality
of the final states will not be affected. Let $\ket{\alpha_{\pm}}$
be two normalized non-orthogonal states that we want to discriminate.
These two initial states do not populate the ancilla level (the third
level). 

The relation between the singular values of $K$ and two-state USD
was studied analytically and graphically in \cite{SVD USD}. The singular
values and the angle between $\ket{\alpha_{\pm}}$ must satisfy $\tan\frac{\phi}{2}=\frac{s_{\text{min}}}{s_{\text{max}}}$
where $\cos\phi=\left|\braket{\alpha_{-}}{\alpha_{+}}\right|$. Since
$s_{\text{max}}=1$ in this problem, we use the result of \cite{SVD USD}
and get:
\begin{eqnarray}
s_{\text{min}} & = & \sqrt{\frac{1-\left|\braket{\alpha_{+}}{\alpha_{-}}\right|}{1+\left|\braket{\alpha_{+}}{\alpha_{-}}\right|}}.
\end{eqnarray}
The weighted laser amplitudes $A_{1},A_{2}$ are given by the first
and second components of the vector $\frac{\ket{\alpha_{+}}+\ket{\alpha_{-}}}{2s_{\text{min}}}$
in the standard basis. After calculating the spectral norm of $H_{\text{RWA}}$
we use (\ref{eq: opt norm action}) to get:
\begin{equation}
T\sqrt{\left|A_{1}\right|^{2}+\left|A_{2}\right|^{2}}=\arcsin\sqrt{\frac{2\left|\braket{\alpha_{+}}{\alpha_{-}}\right|}{1+\left|\braket{\alpha_{+}}{\alpha_{-}}\right|}},\label{eq: atom-laser ineq}
\end{equation}
where $\left|A_{1}\right|^{2}+\left|A_{2}\right|^{2}$ is the optical
power weighted by the dipole coefficients (in larger systems the Hilbert-Schmidt
norm should be used to keep the ``optical power'' interpretation
of the norm). This tradeoff relation between time and effective optical
power ($ $$\left|A_{1}\right|^{2}+\left|A_{2}\right|^{2}$) demonstrates
the main point of this article: the $\text{time}\times\text{energy}$
cost of realizing the discrimination grows when the overlap of the
input states is larger. If the states are orthogonal the RHS of (\ref{eq: atom-laser ineq})
is zero. 

\textbf{\textit{Conclusion$-$}}In this article we have shown that
the unitary embedding of a USD POVM has an intrinsic $\text{time}\times\text{energy}$
cost which depends on the degree of non-orthogonality of the input
states. We have found that the lowest possible embedding cost is obtained
when the diagonal blocks of the unitary are positive. Physically,
this optimal cost is determined by the maximal population transfer
from the system to the ancilla. The optimal cost/resources depends
only on the singular values associated with the desired USD and not
on the size of the ancilla. As shown in the example studied above
this cost has a clear physical significance.%

\end{document}